\documentclass[twocolumn]{aastex62}

\usepackage{amsmath}
\usepackage{amssymb}
\usepackage{hyperref}
\hypersetup{colorlinks,breaklinks,allcolors=blue}
\usepackage{tikz}
\usetikzlibrary{arrows}

\shorttitle{Consequences of Giant Impacts on Early Uranus}
\shortauthors{Kegerreis et al.}

\begin{document}

	\title{\LARGE Consequences of Giant Impacts on Early Uranus for \\
		Rotation, Internal Structure, Debris, and Atmospheric Erosion}

	\correspondingauthor{Jacob A. Kegerreis}
	\email{jacob.kegerreis@durham.ac.uk}
	\author[0000-0001-5383-236X]{J. A. Kegerreis}
	\affiliation{Institute for Computational Cosmology, Durham University, Durham, DH1 3LE, UK}

	\author[0000-0002-8346-0138]{L. F. A. Teodoro}
	\affiliation{BAER/NASA Ames Research Center, Moffett Field, CA, USA}
	\author[0000-0001-5416-8675]{V. R. Eke}
	\affiliation{Institute for Computational Cosmology, Durham University, Durham, DH1 3LE, UK}
	\author[0000-0002-6085-3780]{R. J. Massey}
	\affiliation{Institute for Computational Cosmology, Durham University, Durham, DH1 3LE, UK}
	\author[0000-0001-5646-120X]{D. C. Catling}
	\affiliation{Department of Earth and Space Sciences, University of Washington, Box 351310, Seattle, WA, USA}
	\author[0000-0003-2624-0056]{C. L. Fryer}
	\affiliation{CCS Division, Los Alamos National Laboratory, Los Alamos, NM, USA}
	\author[0000-0003-1950-4789]{D. G. Korycansky}
	\affiliation{CODEP, Department of Earth Sciences, University of California, Santa Cruz, CA 95064, USA}
	\author{M. S. Warren}
	\affiliation{Descartes Labs, 1925 Trinity Drive, Los Alamos, NM, USA}
	\author[0000-0002-2462-4358]{K. J. Zahnle}
	\affiliation{NASA Ames Research Center, Moffett Field, CA, USA}



	\begin{abstract} \noindent
		We perform a suite of smoothed particle hydrodynamics simulations
		to investigate in detail the results of a giant impact on the young Uranus.
		We study the internal structure, rotation rate,
		and atmospheric retention of the post-impact planet,
		as well as the composition of material ejected into orbit.
		Most of the material from the impactor's rocky core falls in to the core of the target.
		However, for higher angular momentum impacts,
		significant amounts become embedded anisotropically as lumps in the ice layer.
		Furthermore, most of the impactor's ice and energy is deposited
		in a hot, high-entropy shell at a radius of $\sim$3~$R_\oplus$.
		This could explain Uranus' observed lack of heat flow from the interior
		and be relevant for understanding its asymmetric magnetic field.
		We verify the results from the single previous study of lower resolution simulations
		that an impactor with a mass of at least 2~$M_\oplus$ can
		produce sufficiently rapid rotation in the post-impact Uranus
		for a range of angular momenta.
		At least 90\% of the atmosphere remains bound to the final planet after the collision,
		but over half can be ejected beyond the Roche radius by a 2 or 3~$M_\oplus$ impactor.
		This atmospheric erosion peaks for intermediate impactor angular momenta
		($\sim$$3\times10^{36}$~kg~m$^2$~s$^{-1}$).
		Rock is more efficiently placed into orbit
		and made available for satellite formation
		by 2~$M_\oplus$ impactors than 3~$M_\oplus$ ones,
		because it requires tidal disruption that is suppressed by the more massive impactors.
	\end{abstract}



\section{Introduction} \label{sec:intro}

Uranus spins on its side.
With an obliquity of 98$^\circ$
and its major moons orbiting in the same tilted plane,
the common explanation is that a giant impact
sent the young Uranus spinning in this new direction
\citep{Safronov1966}.
This impact might also help explain other phenomena,
such as the striking differences between Uranus' and Neptune's satellite systems \citep{Morbidelli+2012,Parisi+2008},
the remarkable lack of heat from Uranus' interior
\citep{Stevenson1986,Podolak+Helled2012,Nettelmann+2016},
and its highly asymmetrical and off-axis magnetic field \citep{Ness+1986}.
Until now, this violent event itself has been little studied since the first
smoothed particle hydrodynamics (SPH) simulations of \citet{Slattery+1992}.

Uranus' equatorial ring and satellite system is remarkable in several respects.
It features a set of regular, prograde, major moons,
a compact inner system of rings and small satellites,
and a distant group of irregular moons.
The inner system and major moons are hypothesised to have formed either
from a post-impact debris disk \citep{Stevenson1986,Slattery+1992}
or from a pre-impact proto-satellite disk
that was destabilised by the post-impact debris disk
and rotated to become equatorial \citep{Morbidelli+2012,Canup+Ward2006}.
The more-distant irregular satellites are thought to have been captured
after the impact \citep{Parisi+2008}.

The interior structure of Uranus is poorly understood. Surface
emission is in approximate equilibrium with solar insolation, implying
that negligible heat flows out from the planet,
in striking contrast with the other giant planets \citep{Pearl+1990}.
This might be explained by restricted interior convection, perhaps caused
by the deposition of the impactor's energy into a thin shell
\citep{Stevenson1986,Podolak+Helled2012}. Such a thermal boundary
layer between an outer H--He-rich envelope and an inner ice-rich layer
was the crucial ingredient for the evolutionary model of Uranus produced
by \citet{Nettelmann+2016} that was consistent with both
heat flow and gravitational moment measurements.

In contrast with terrestrial planets,
the magnetic field of Uranus measured by {\it Voyager} 2 was not
dominated by the dipole component. Higher order moments contributed
significantly, and the dipole itself was both offset by approximately
0.3 Uranus radii from the centre of the planet and tilted by $60^\circ$
relative to Uranus' rotation axis \citep{Ness+1986}. Dynamo models producing
similar magnetic fields have been constructed using a layer of convecting
electrically conducting ices
\citep{Stanley+Bloxham2004,Stanley+Bloxham2006,Soderlund+2013}. A feature of
some of these models is the presence of a stably stratified fluid
layer interior to the zone where the magnetic field is generated.

As a separate source of motivation,
while the ice giants Uranus and Neptune do not receive as much
attention as the nearer bodies in the solar system, they represent the
closest analogues to the mini-Neptune-class exoplanets that are the most
frequently discovered by Kepler \citep{Batalha2014}. Given the
detection efficiencies, these planets are typically found on orbits
with periods of the order of 100 days \citep{Fressin+2013}, but
have nevertheless stimulated attempts to understand the atmospheres and histories of
our ice giants in order to provide context for these exoplanet
observations \citep{Fortney+2013}.

The first simulations of a giant impact onto a proto-Uranus,
albeit in one dimension, were done specifically
to investigate whether the shock from the collision
would blast away Uranus' hydrogen--helium atmosphere \citep{Korycansky+1990}.
This gas has a much lower mass fraction and density than the inner ice and rock material,
so requires high resolution to simulate. For this reason,
\citet{Korycansky+1990} restricted their study to a one-dimensional
spherically symmetric model where the impactor mass and some of its energy was injected into
the proto-Uranus core, and the remaining energy was placed into the
atmosphere. The retained atmospheric mass was found to depend
sensitively upon the amount of energy deposited directly into the
atmosphere, offering the possibility that the presence of Uranus'
current atmosphere might constrain allowable impact scenarios.

Building on the pioneering work of \citet{Benz+1986}, who used
SPH simulations to model the Moon-forming giant impact on the Earth,
\citet{Slattery+1992} (hereafter S92) produced, to our knowledge,
the only paper to date with three-dimensional
hydrodynamical simulations of the hypothesised
impact event that befell the proto-Uranus.
While the $<$$10^4$-particle SPH simulations of S92 did
not resolve the atmosphere, they studied collisions between a
1 and 3~$M_\oplus$ differentiated impactor containing iron, dunite, and
ice and a similarly differentiated proto-Uranus with hydrogen and
helium mixed into its ice layer.
For impactor masses above 1~$M_\oplus$, they found a wide
range of impact scenarios that led to a sufficiently rapidly spinning
planet. Most of these collisions left ice in orbit, but only the
higher angular momentum ones also placed any rock or iron into orbit, as
might be expected if this material is subsequently to form any of the
currently observed regular moons. Uranus' satellites comprise
only $\sim$10$^{-4}$ of the total system mass---%
the same mass fraction as the other giant planets---%
corresponding to just less than the mass of a single particle in S92's simulations.

In this paper, we present new simulations of the impact
with orders of magnitude better mass resolutions than those of S92,
allowing the detailed modelling of, for example,
Uranus' atmosphere and its fate;
the deposition of the impactor's material and energy inside Uranus;
the post-impact debris disk, in particular the amount, distribution,
and composition of material available for satellite formation;
and the testing of S92's original conclusions for
the types of impacts that could have produced the present-day spin.

Section~\ref{sec:methods} describes the methods used to construct initial
conditions and run the impact simulations. Our results are reported
and discussed in section~\ref{sec:results}, and the findings are
summarised in section~\ref{sec:conclusions}.

\section{Methods}\label{sec:methods}

In this section, we first outline the equations of state (EoS) used for the
various materials in the simulations
and the generation of the initial conditions,
followed by detailing the simulation runs themselves.

\subsection{EoS and Initial Conditions}\label{sec:eos}

Planets contain multiple and complex materials,
so a few different EoS---%
which relate the pressure, density,
and temperature or specific internal energy---%
need to be specified for our SPH simulations.

Our proto-Uranus contains a rocky core
(SiO$_2$, MgO, FeS, and FeO),
icy mantle (H$_2$O, NH$_3$, and CH$_4$),
and atmosphere with a solar composition mix of hydrogen and helium.
These materials were used for the Uranus model of
\citet{Hubbard+MacFarlane1980} (hereafter HM80), and for our
simulations described here we use the EoS as presented in their
paper (appendix \ref{sec:app:eos}).
These relatively straightforward EoS provide us with some baseline
simulation results that will in the future be compared with more advanced
EoS, such as those more recently determined for ices and hydrogen and helium
\citep{Nettelmann+2008,Redmer+2011,Militzer+Hubbard2013,
	Bethkenhagen+2013,Wilson+2013,Bethkenhagen+2017}.

We use a range of impactor masses of $M_{\rm{i}}=1$, 2, and 3~$M_\oplus$ and,
under the assumption that little mass escapes during the impact,
set the mass of the proto-Uranus to be 14.536~$M_\oplus-M_{\rm{i}}$.
The proto-Uranus is differentiated into the three distinct layers
described above.
The impactor is given no atmosphere,
so it has only a rocky core surrounded by an icy mantle,
with the ice/rock mass ratio matching that in the proto-Uranus.

To determine the amounts of rock, ice, and atmosphere in the two bodies,
we first create a spherically symmetric three-layer model for the present-day
Uranus, assuming hydrostatic equilibrium. The assumed outer
boundary conditions are a pressure of 1 bar and a temperature of
60 K at a radius of 3.98~$R_\oplus$.
We then iterate the radii of the layer boundaries until
the profile contains the desired total mass (14.536~$M_\oplus$) and
a reduced moment of inertia of $I/(MR^2)=0.21$.
The outer temperature is slightly lower than the measured value (75 K),
in order that this simple model can approach the observed reduced
moment of inertia of 0.22 \citep{Podolak+Helled2012}.
We find an ice-rich body,
with 2.02, 11.68, and 0.84~$M_\oplus$
in the rock, ice, and atmosphere layers, respectively,
with inner boundaries at radii of 1.0 and 3.1~$R_\oplus$.
There is considerable uncertainty in the composition of Uranus;
this ratio of ice to rock is comparable with that in the model of \citet{Nettelmann+2013},
but larger than that found by HM80
and almost twice the solar system value adopted by S92.

\begin{figure}[t]
	\centering
	\plotone{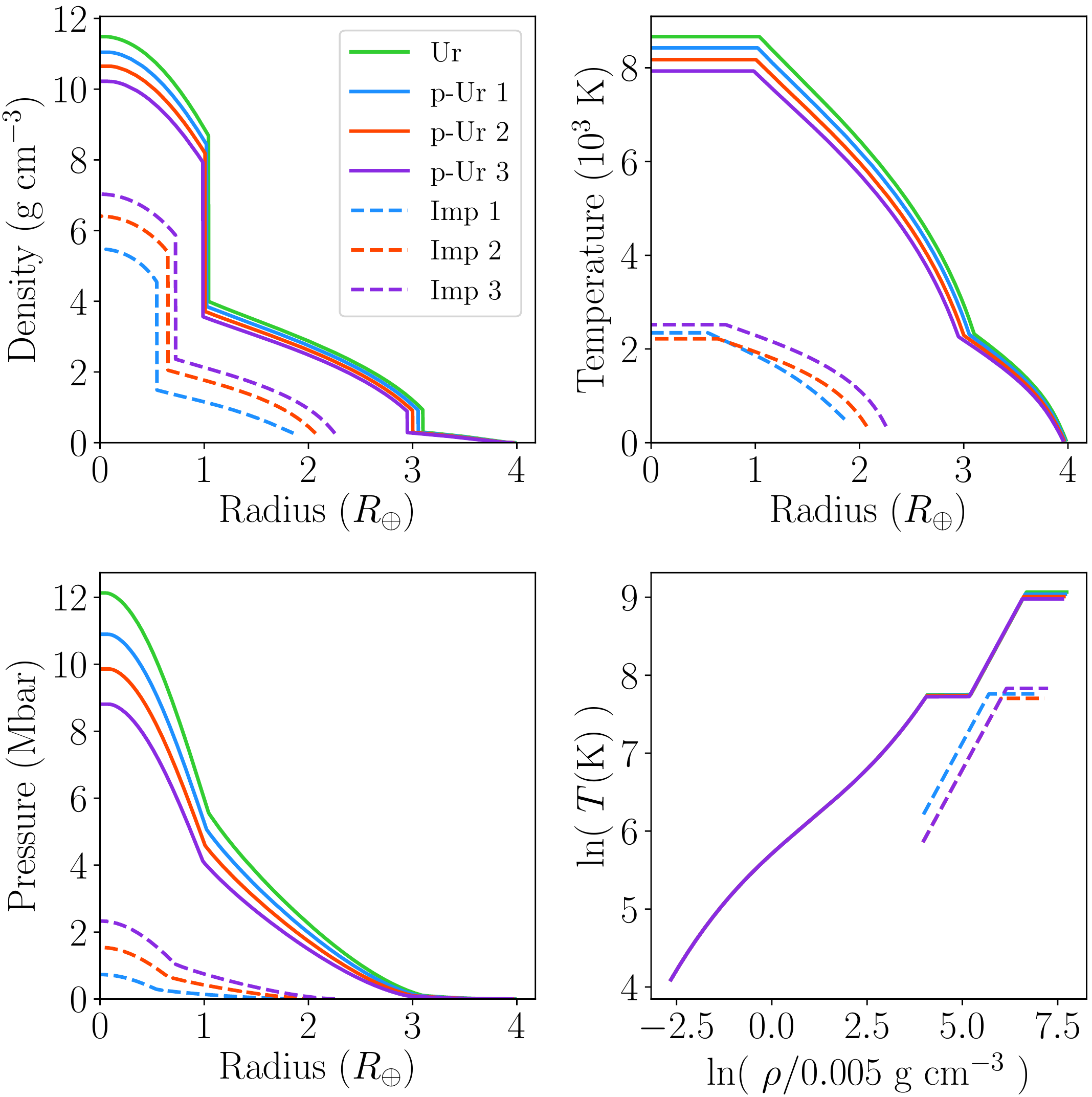}
	\caption{The density, temperature, and pressure profiles of our
		Uranus (Ur) model and the three pairs of
		proto-Uranus (p-Ur) and impactor (Imp) bodies.
		The bottom-right panel shows the temperature--density relations
		assumed in the various objects.
		The colours	correspond to different masses of the impactor
		as labelled in the legend (in units of $M_\oplus$).
		The green line shows the model Uranus
		whose mass we split into the proto-Uranus and impactor.
		\label{fig:init_profs}}
\end{figure}

The density, temperature, and pressure profiles for our Uranus model as
well as the three proto-Uranus and impactor pairs are shown in Figure~\ref{fig:init_profs}.
Also included are the density--temperature relations,
showing our isothermal rocky cores,
the approximately adiabatic power-law relation for the ice mixture used by HM80,
and their fitted polynomial adiabat for the atmosphere.

One simplification present in our initial conditions is
the lack of compositional mixing between the different layers.
For instance, S92 included H--He mixed into the icy mantle of their
proto-Uranus, and the model of \citet{Hubbard+Marley1989} had some ice
mixed into the rocky core. Given the uncertainties in the current
internal structure of Uranus, and the much larger uncertainties in those of
the proto-Uranus and impactor, we opt for simply differentiated
bodies for these initial investigations.

The impacts we consider are violent enough to
dominate over any pre-existing rotation,
so our proto-Uranus (and impactor) begins without any spin.
This spherical symmetry also makes the generation of initial conditions much simpler,
so we leave investigating the effects of pre-impact spin for a future study.

\begin{figure*}[t]
	\centering
	\plotone{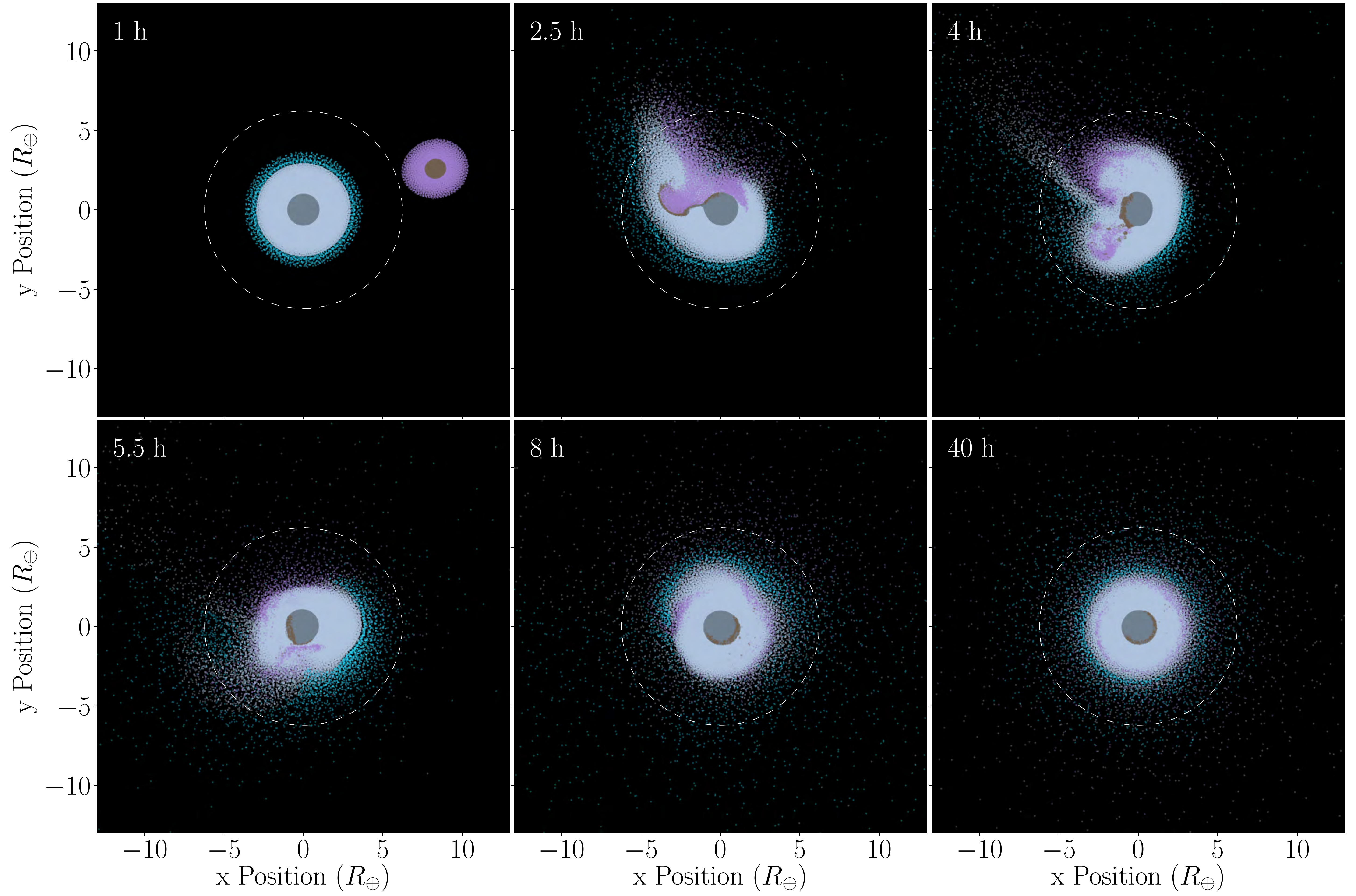}
	\caption{Snapshots from a low angular momentum impact
		simulation with a 2~$M_\oplus$ impactor and
		$L=2\times10^{36}$~kg~m$^2$~s$^{-1}$.
		Particles between $z=0$ and $-13$~$R_\oplus$ are shown,
		coloured by material type and originating body.
		Light and dark grey show the target's ice and rock material, respectively,
		and purple and brown show the same for the impactor.
		Light blue is the target's atmosphere.
		The white dashed circle traces out the current Roche radius of Uranus for reference.
		The snapshot times are given to the nearest half hour since the start of the simulation.
		This figure is available as an animation in the online version.
		\label{fig:m2L2v5}}
\end{figure*}
\begin{figure*}[t]
	\centering
	\plotone{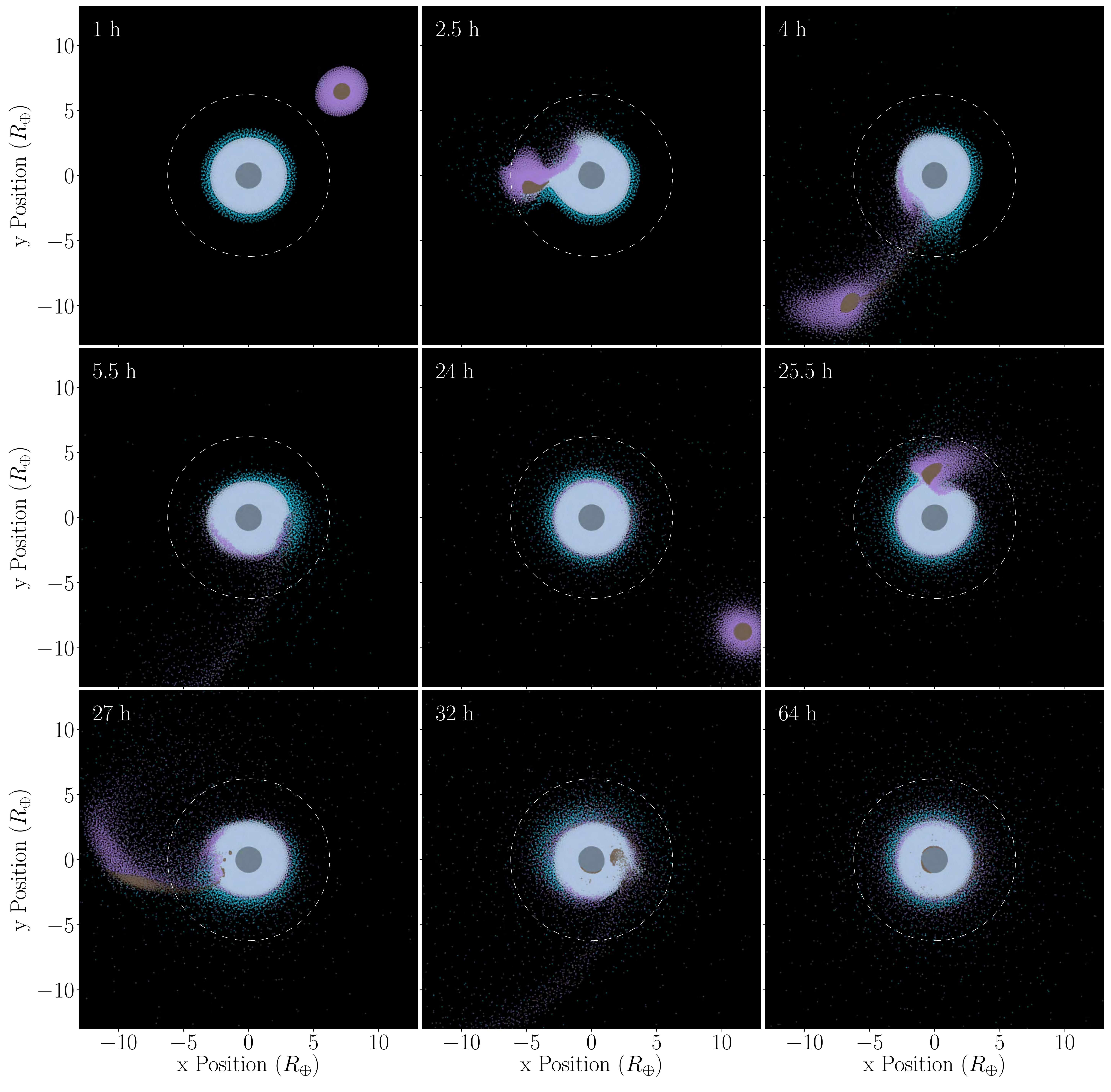}
	\caption{As for Figure~\ref{fig:m2L2v5}, but for a high
		angular momentum impact simulation with a 2~$M_\oplus$ impactor and
		$L=5\times10^{36}$~kg~m$^2$~s$^{-1}$.
		This figure is available as an animation in the online version.
		\label{fig:m2L5v5}}
\end{figure*}

\subsection{Particle Placement}

The EoS for the ice and rock materials being simulated are very stiff,
i.e. a small variation in density changes the pressure dramatically.
It is therefore important to reduce particle noise
when sampling the desired mass distribution with particles.
Even small deviations from the profile density will lead to transient
behaviour that can take a long time to settle,
during which the particle distribution may also significantly change.

We have developed a new algorithm for quickly creating low-noise particle distributions
for an arbitrary spherically symmetric mass distribution,
such that every particle's SPH density is within 1\% of the desired value
(Kegerreis et al., 2018, \emph{in prep.}).
The code is publicly available at
\href{https://github.com/jkeger/seagen}{https://github.com/jkeger/seagen}
and the \verb|seagen| python module can be installed directly with pip.

\citet{Raskin+Owen2016} and \citet{Reinhardt+Stadel2017} developed
comparable approaches to the challenge of placing particles to represent
spherically symmetric mass distributions in ways that avoided the
various problems of lattice-based methods \citep{Herant1994}. However,
we found that the method of \citet{Raskin+Owen2016} leads to
a few particles in every shell having significant overdensities,
causing unrealistically high pressures with the stiff EoS.
The approach of \citet{Reinhardt+Stadel2017} cannot place arbitrary
numbers of particles in each shell. Consequently, some particles show
SPH densities more than 5\% discrepant from the profile.

Our method leads to initial conditions that are
close to equilibrium and quick to produce,
avoiding the need for a lengthy simulation to
relax the system. Briefly, our method involves distributing any
arbitrary number of particles in spherical shells
(a nontrivial problem \citep{Saff+Kuijlaars1997}),
starting by dividing a spherical shell into equal-area regions
arranged into iso-latitude bands.
An empirical stretch away from the poles is then
applied so that the particles, when placed in the centres of
these regions, all have very similar densities as determined using the
relevant SPH smoothing kernel.
All particles have a similar mass,
with mass variations of $\sim$3\%
because of the integer numbers in each shell.
Concentric shells can then be set up to follow precisely an
arbitrary radial density profile with very low scatter in each shell.

The small density discrepancies in this particle placement scheme
result in average transient particle speeds
in our initial conditions that are already under 1\% of the escape speed.
A quick relaxation simulation, described in section~\ref{sec:SPH},
further reduces this by an order of magnitude.

\subsection{SPH Simulations} \label{sec:SPH}

All simulations were run with a version of
the parallel tree-code \texttt{HOT} \citep{Warren+Salmon1993}
that has been modified to include SPH \citep{Lucy1977,Gingold+Monaghan1977}
and the relevant EoS described in section~\ref{sec:eos}
(see also appendix~\ref{sec:app:eos}).
The SPH formulation is described in \citet{Fryer+2006}.
Particles with different EoS are adjacent at the boundaries,
which can cause problems in SPH given the sharp density changes,
in addition to the known issues regarding the mixing of materials
\citep{Woolfson2007,Hosono+2016,Deng+2017}.
To verify the stability of our model planets
given the lack of any special boundary treatments in this simple SPH formulation,
we ran a simulation where the impactor misses the target
but is slightly tidally disrupted,
so that any problems would not be hidden in the middle of a violent impact.
We confirmed that the pressure at the core-mantle boundary
evolved smoothly and remained stable,
showing the same `unloading' behaviour tested by \citet[][Fig.~2(b)]{Asphaug+2006}.

Initial simulations of the proto-Uranus and impactor
for 10,000~s in isolation were performed
including a damping force to further reduce
any remaining small fluctuations in density.
At the end of these simulations,
the total kinetic energy was decreased from a fraction of
$\sim$$10^{-5}$ to below $10^{-6}$ of the total energy.
This corresponds to reducing the maximum particle velocity
to below 1\% of the target planet's escape speed,
with an average random velocity of $\sim$0.1\% of the escape speed.

\begin{figure*}[t]
	\centering
	\plotone{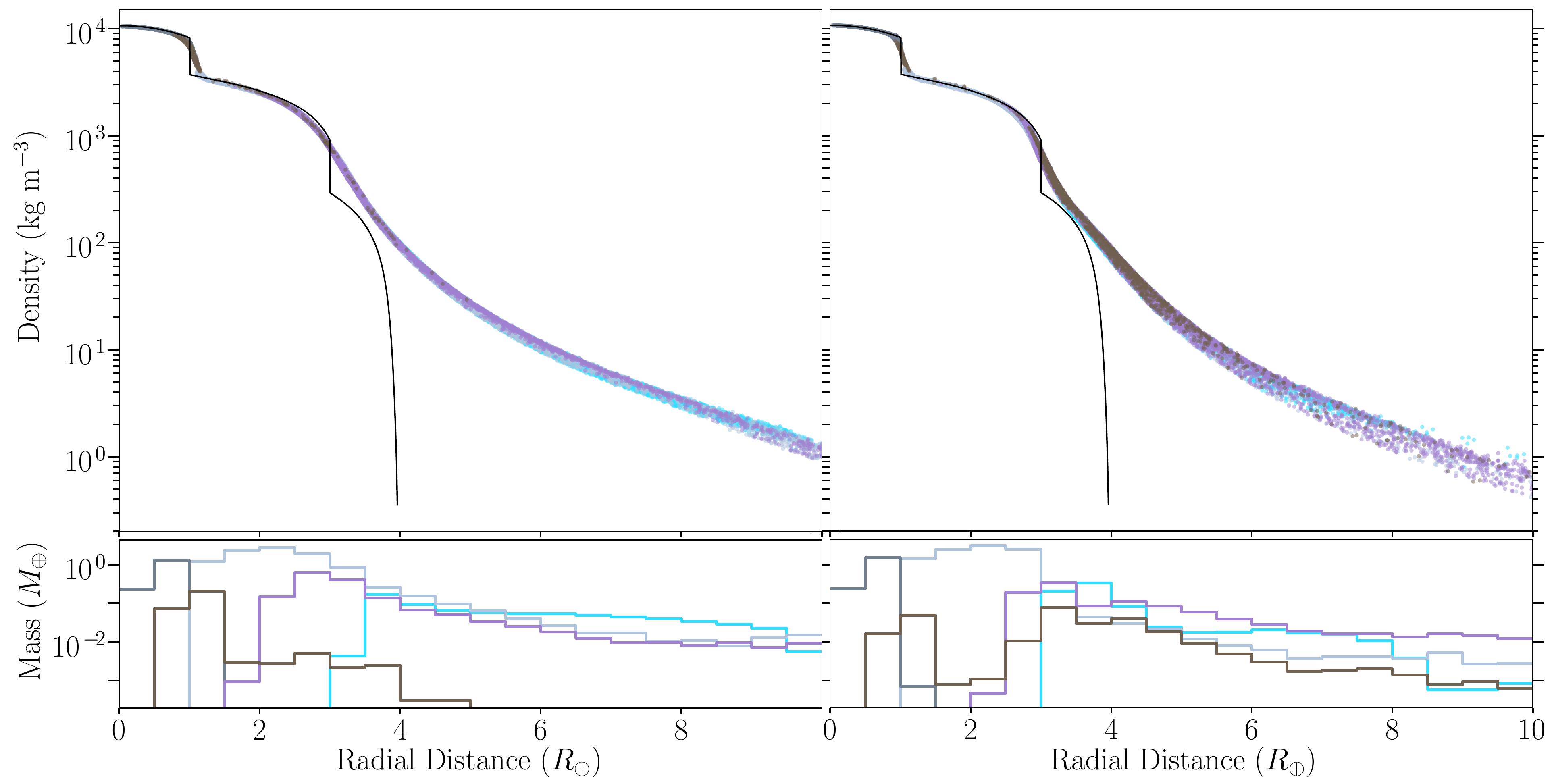}
	\caption{The final radial density profiles for the same relatively
		head-on (left) and grazing (right) impacts
		as in Figures~\ref{fig:m2L2v5} and~\ref{fig:m2L5v5}.
		The black line shows the proto-Uranus density profile.
		The lower panels show the mass of particles
		in radial bins of width 0.5~$R_\oplus$,
		split by material type and originating body.
		Light and dark grey show the target's ice and rock material, respectively,
		and purple and brown show the same for the impactor.
		Light blue is the target's atmosphere.
		\label{fig:rho_profs}}
\end{figure*}

Prior to impact, the impactor and proto-Uranus both become
distorted by the gravitational tides from the other
object. The subsequent evolution can depend significantly upon these
departures from sphericity at impact. Thus, for an accurate
reproduction of the collision, it is necessary to start the impactor
sufficiently far enough away that these tidal distortions are faithfully
followed. To achieve this, we placed the impactor such that its closest
particle to the proto-Uranus received a 10 times larger gravitational
force from the rest of the impactor than from the proto-Uranus. This
amounts to separations of $\sim$22, 16, and 14~$R_\oplus$
for the 1, 2, and 3~$M_\oplus$ impactors respectively
(appendix~\ref{sec:app:init_cond}).

Separate suites of impacts were created with just over $10^5$ and
$10^6$ particles to test the resolution-dependence of our results.
The angular momenta of the systems ranged from 1 to $10 \times
10^{36}$~kg~m$^2$~s$^{-1}$. This was achieved by changing the impact
parameter, while keeping the relative velocity at infinity
fixed at 5~km~s$^{-1}$, following S92 (appendix~\ref{sec:app:init_cond}).
Three head-on impacts were also simulated,
one for each impactor mass.
These of course cannot produce the required spin but
are useful comparisons for investigating the other consequences of a collision.
A set of otherwise-identical simulations
with velocities at infinity ranging from 1 to 9~km~s$^{-1}$ were also performed
to confirm that this choice does not significantly affect the results.

Depending on the angular momentum and impactor mass, the time taken
for the impact to complete and leave a settled planet varied from
roughly 1 to 7 Earth days.
The simulations were stopped once the results presented in this paper
were not changing over timescales of 10,000~s.
Using a Courant factor of 0.3 gave typical simulation timesteps
of 5--10~s and 2.5--5~s for the $10^5$ and $10^6$ particle runs, respectively,
meaning that the impact simulations typically contained $\sim$$10^5$ steps.

\newpage
\section{Results}\label{sec:results}

The results of the simulations are described in this section,
starting with a broad description of the post-impact distribution of material.
This enables us to define three mutually exclusive categories into which
the particles are placed: `planet', `orbit' and `unbound'.
We then describe in more detail the properties of the planets
that are produced, before turning our attention to the composition of
the orbiting debris cloud exterior to the Roche radius and the
fraction of the H--He atmosphere that is retained within the Roche
radius after the impact.

Given the large number of simulations,
we will focus, in particular, on two 2~$M_\oplus$-impactor simulations with
low ($L=2\times10^{36}$~kg~m$^2$~s$^{-1}$)
and  high ($L=5\times10^{36}$~kg~m$^2$~s$^{-1}$) angular momenta,
as archetypal examples of $\sim$head-on and grazing impacts respectively.
Figures~\ref{fig:m2L2v5} and~\ref{fig:m2L5v5} show snapshots
from these two giant impact simulations,
included as animations in the online version.
These illustrate the typical features of all the impacts,
with most of the impactor's rock ending up on the edge of the
core of the final planet, while the impactor's ice is deposited into the
outer regions of the icy mantle. At higher angular momenta,
multiple passes and tidal stripping of the impactor
leave more material in orbit around the final planet.
Full animations of the impacts are also available to view at
\href{http://icc.dur.ac.uk/giant_impacts}{icc.dur.ac.uk/giant\_impacts}.

\newpage
\subsection{Material Distribution}\label{sec:material_dist}

The density profiles of the final mass distributions in the example low and
high angular momentum impacts are shown in
Figure~\ref{fig:rho_profs}. For the more head-on collision, the impactor
core is delivered more efficiently to the core of the final
planet. This type of collision also places slightly more impactor ice deeper
into the final planet than the relatively grazing impact. As a
consequence, more of the proto-Uranus' ice and atmosphere is
jettisoned into orbit around the final planet or ejected from the
system entirely.

\begin{figure}[t]
	\centering
	\plotone{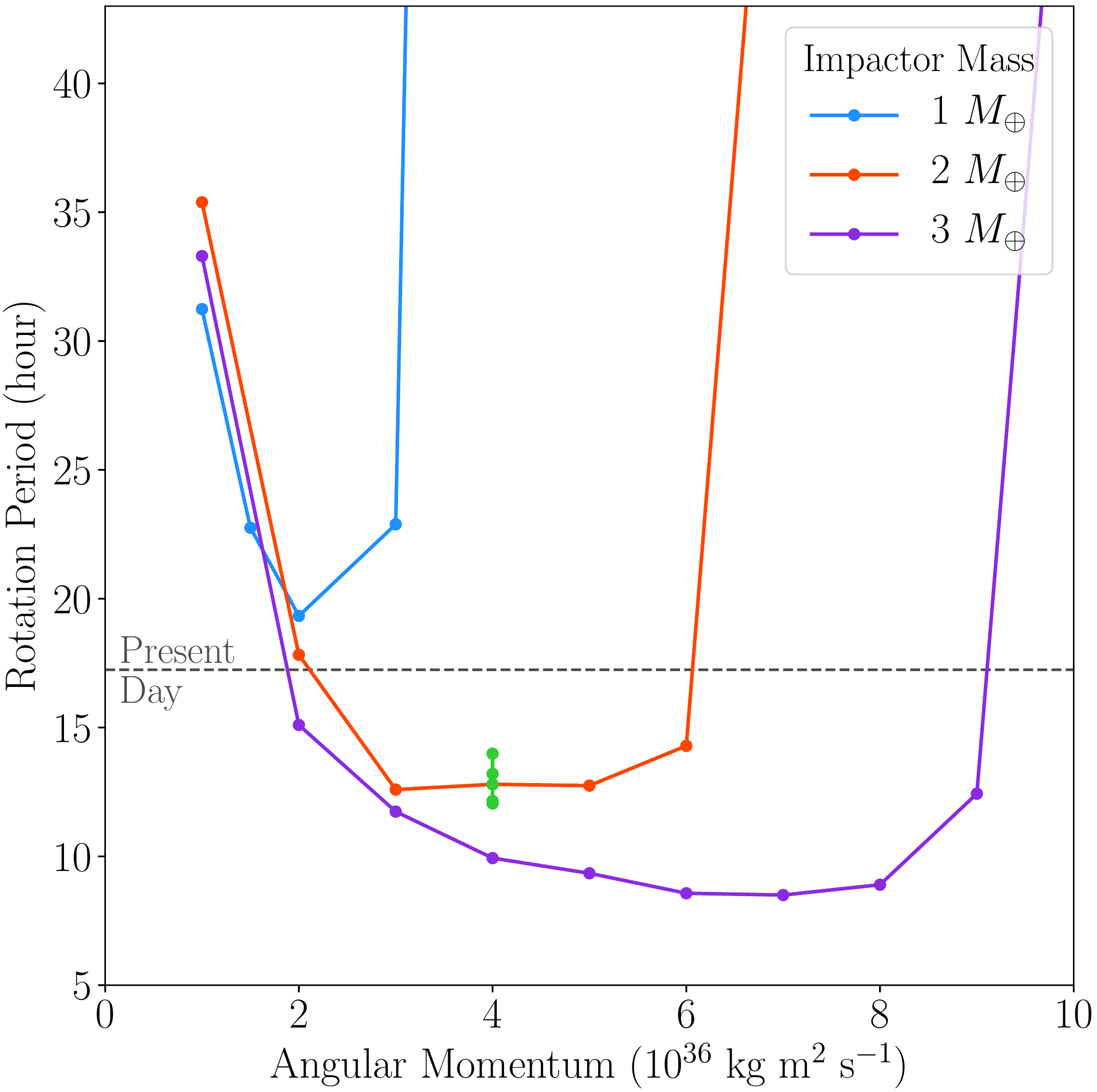}
	\caption{Median rotation periods for particles in the final
		planets produced by runs with different angular momenta
		and impactor masses, as given in the legend.
		The rotation period of each particle is calculated
		from its tangential velocity and distance from the $z$ axis.
		All planet particles have negligible velocities in the radial and $z$ directions.
		The green points show the 2~$M_\oplus$-impactor simulations with
		velocities at infinity of 1--9~km~s$^{-1}$ instead of the default 5~km~s$^{-1}$.
		The dashed horizontal line shows the current rotation rate of Uranus of
		17.24~hr \citep{Warwick+1986}.
		\label{fig:rotation_periods}}
\end{figure}
\begin{figure}[t]
	\centering
	\plotone{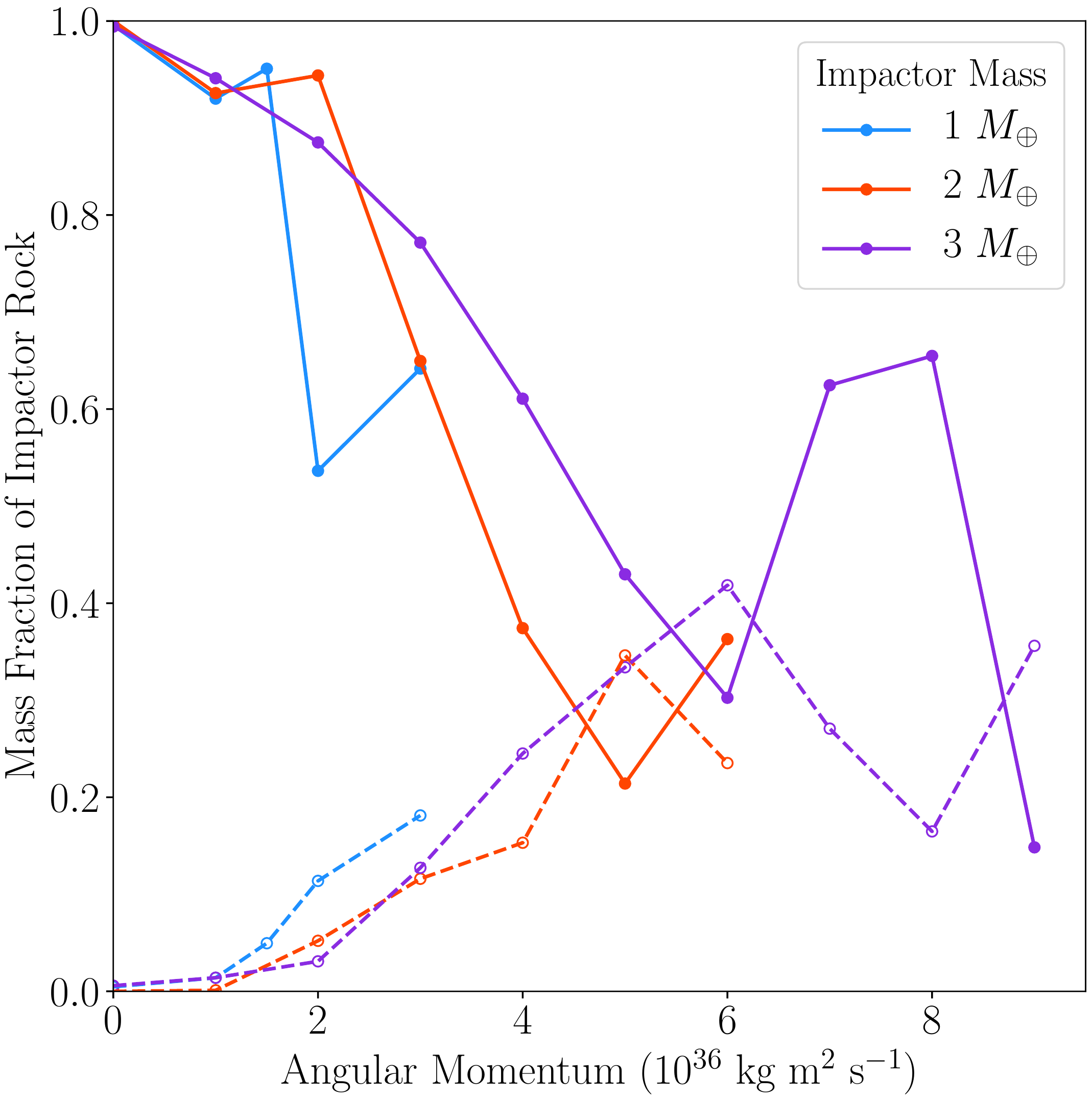}
	\caption{The fraction of impactor rock reaching the core of
		the final planet ($<1.3$~$R_\oplus$, solid lines) or deposited
		elsewhere in the planet (dashed lines) as a function of
		impactor mass (as given in the legend) and angular
		momentum.
		\label{fig:imp_roc_core}}
\end{figure}

The smooth decrease in density seen for both cases in
Figure~\ref{fig:rho_profs} raises the question of how to define
the edge of the final planet, which is also slightly flattened due to the
rotation that it has acquired. We choose to do this using a
friends-of-friends (FoF) algorithm \citep{Davis+1985}. This links
together particle pairs that are separated by less than some user-defined distance
and effectively finds groups of linked particles bounded by an isodensity
surface. Using a linking length of 0.3~$R_\oplus$ for the low resolution
simulations, and scaling by the inverse cube root of the particle
number for the high resolution cases leads to
a final planet with a radius of $\sim$4~$R_\oplus$
and a mass that is insensitive to small changes of the linking length.

A significant amount of material external to this planet is,
nevertheless, gravitationally bound to it.
We will refer to this as orbiting material. The remaining mass
is unbound. The orbiting material can be further divided into that
within the Roche radius, which one would expect to accrete relatively
quickly onto the planet, and that outside this radius, which is
available to form moons. While our simulated planets have Roche radii
of 5.5--5.8~$R_\oplus$ (for a satellite density of 1~g~cm$^{-3}$),
the Roche radius of present-day Uranus is 6.2~$R_\oplus$. When considering the
material available for moon formation and the distribution of the
post-impact H--He, we will use radii of $6\pm0.5$~$R_\oplus$ to allow for the
uncertainty in the planet's mass and choice of satellite density.

\subsection{Resulting Planet}

With the final planets defined as described in section~\ref{sec:material_dist},
we can study their rotation rates and internal structures.
These properties are discussed in the following two subsections.

\subsubsection{Rotation Rate}

Figure~\ref{fig:rotation_periods} shows how the rotation period varies with
impactor mass and angular momentum.
Despite using different
proto-Uranus and impactor models from those of S92,
we find broadly similar results.
There is no 1~$M_\oplus$ impactor with a
relative velocity at infinity of 5~km~s$^{-1}$ that can
produce a sufficiently rapidly rotating planet.
Both 2 and 3~$M_\oplus$ impactors are able to satisfy this requirement,
provided that the impactor is bringing an angular momentum of at least
$2\times10^{36}$~kg~m$^2$~s$^{-1}$.
At first, raising the angular momentum increases the final spin.
However, for very high angular momentum values, to the right of the figure,
the impactor starts to only graze and eventually misses the target,
making it unable to transfer enough of its huge angular momentum.

Our range of simulation numbers of particles
shows that these results vary little with numerical resolution,
and find them to be already well-determined with the low number of particles adopted by S92.
So, the general agreement of (and any differences between)
our rotation-rate results and theirs
is primarily testing the different models for the colliding bodies and
the materials within them, rather than showing numerical effects.

\subsubsection{Interior}

\begin{figure}[t]
	\centering
	\plotone{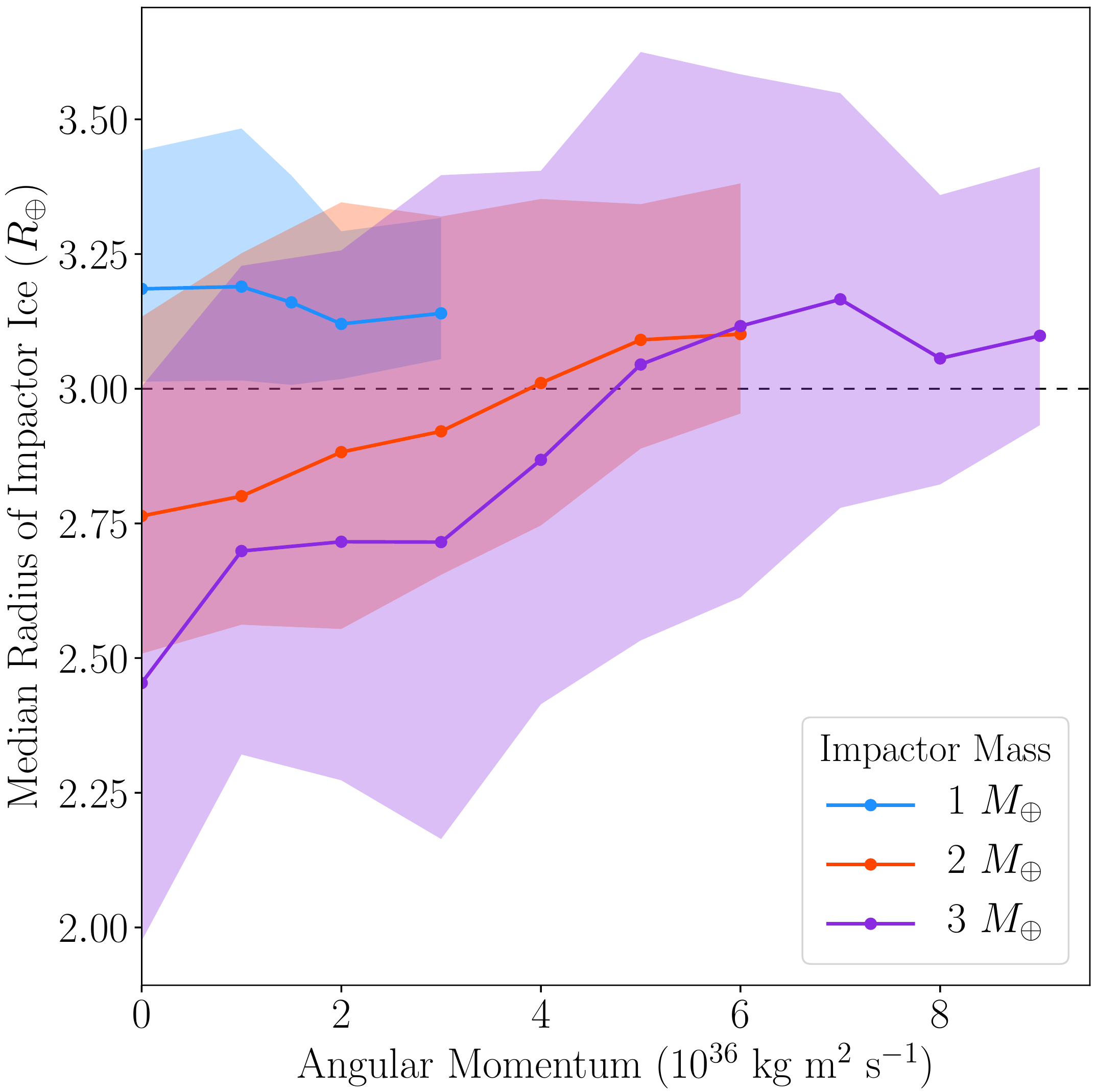}
	\caption{The radius of deposition of the impactor ice
		as a function of impactor mass and angular momentum.
		Shaded regions show the 1-$\sigma$ percentile range
		of the radius distributions.
		The dashed line shows the approximate radius
		of the ice-atmosphere boundary in the proto-Uranus targets.
		\label{fig:imp_ice_radius}}
\end{figure}
\begin{figure*}[t]
	\centering
	\plotone{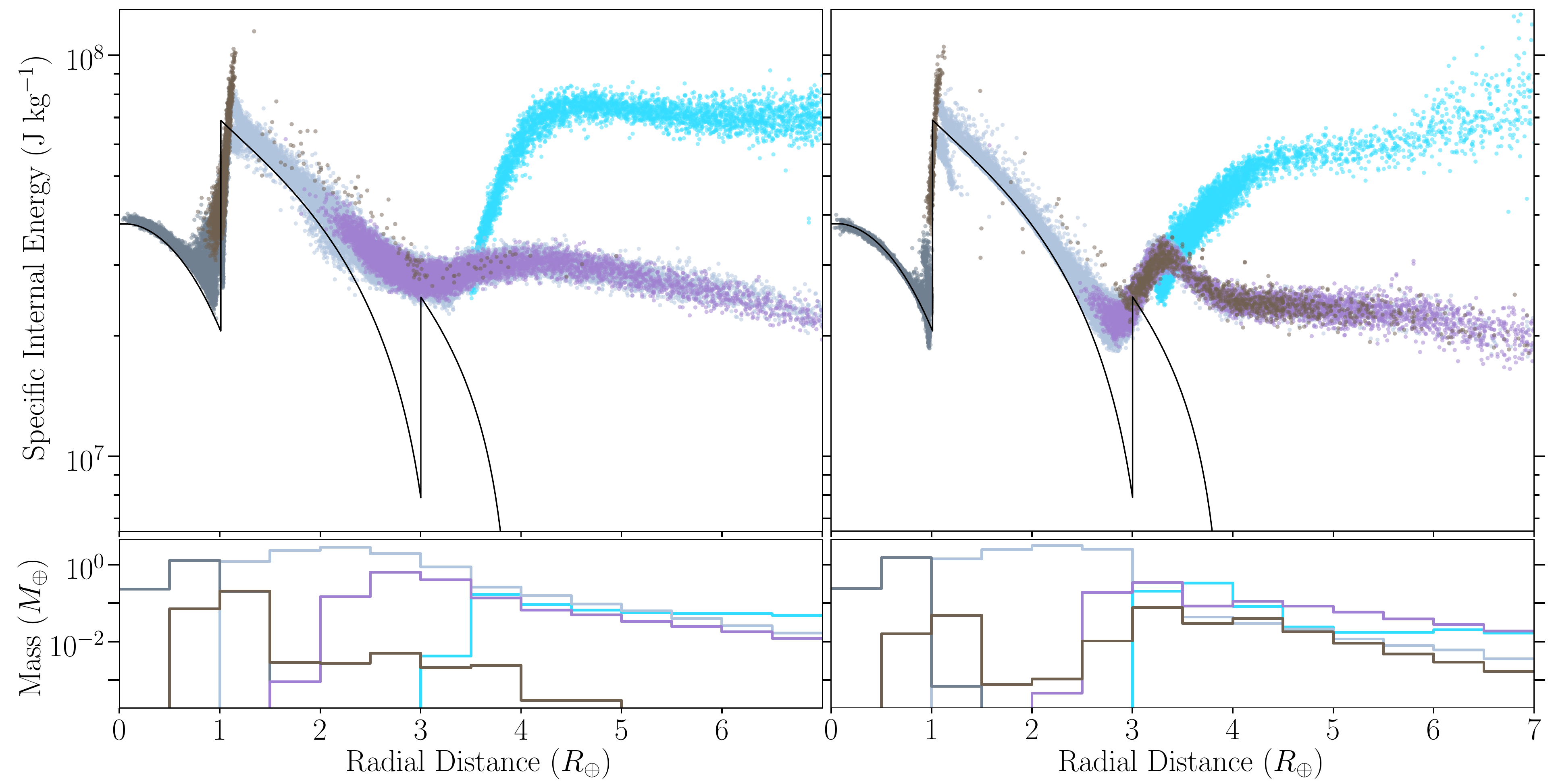}
	\caption{The final radial internal energy profiles for the same relatively
		head-on (left) and grazing (right) impacts
		and the same lower-panel histograms
		of mass per radial bin as in Figure~\ref{fig:rho_profs}.
		\label{fig:u_profs}}
\end{figure*}

The density profiles within the planet and their decomposition into
material types from the two colliding bodies are shown for the low
and high angular momentum impacts in
Figure~\ref{fig:rho_profs}. Considering the suite of simulations in full,
Figures~\ref{fig:imp_roc_core} and~\ref{fig:imp_ice_radius} show the
destinations of the impactor rock and ice within the planet
respectively.

It is apparent from Figure~\ref{fig:imp_roc_core} that the head-on
collisions deliver practically all of their impactor rock to the core.
However, as the angular momentum is raised,
the fraction of the rock in the impactor that is deposited
higher up in the ice layer of the final planet
or even into orbit increases significantly.
The non-monotonic behaviour at high angular momenta is a consequence of
an initially grazing impact sometimes leading to a much more head-on
secondary collision of the core after the ice has been stripped
and some angular momentum lost.
Up to 40\% of the rock in 2 and 3~$M_\oplus$ impactors
can be left embedded in the icy mantle for
sufficiently high angular momentum collisions.
In our $\sim$$10^6$-particle simulations,
this rock is present in well-resolved, mostly spherical lumps.
Such inhomogeneities will be investigated in detail with higher resolution
simulations in the future, but this is beyond the scope of this initial study.

The rock that is added during the collision is generally not distributed
isotropically with respect to the centre of the planet. For the
2~$M_\oplus$ impactors, 90\% of the delivered rock covers
only $\sim$50\% of the $4\pi$ steradians subtended at the planet's centre.
This increases to $\sim$70\% coverage for the 3~$M_\oplus$ impactors. The ice
that is deposited tends to be more isotropically distributed than the rock, unless
the impact is head on in which case 90\% of the delivered ice
subtends only $\sim$40\%$\times4\pi$ steradians,
independent of impactor mass.

Where this impactor ice is deposited may have profound implications for
the current internal structure of and heat flow from
Uranus. Figure~\ref{fig:imp_ice_radius} shows the final destinations in radius
of the impactor ice. For the 1~$M_\oplus$ impactors,
the ice is mostly deposited on top of the pre-existing icy mantle,
independently of angular momentum, because the impactor is not massive enough to
sufficiently disturb the proto-Uranus. However, the larger projectiles are able to
inject ice deeper into the final planet, particularly for the lower
angular momentum collisions. These more head-on collisions also lead
to a slightly thicker zone that is infiltrated by impactor ice
(interquartile range spanning $\sim$1~$R_\oplus$) than the higher angular
momentum cases, which do not penetrate as significantly into the mantle
and can spread the impactor ice out into a thinner layer.

In addition to delivering mass, the impactor deposits a significant
amount of energy into the final planet.
The radial profiles of specific internal energy out to a little beyond the Roche
radius are shown in Figure~\ref{fig:u_profs},
as well as the initial profile with its $\sim$adiabatic ice layer.
For both low and high angular momentum collisions, the impactor rock that
reaches the edge of the final planet's core is much hotter
than the largely undisturbed proto-Uranus rock.
In high angular momentum collisions,
a similar temperature inversion is created
near the boundary between the ice and atmosphere,
where the impactor ice has been delivered,
creating a high-entropy layer of hot material.
This sub-adiabatic energy gradient is also present in the icy mantle
following low angular momentum collisions,
but it is less dramatic because of the broader range of radii
into which the impactor mass and energy has been deposited.

Investigating the extent and implications of this departure
from adiabatic behaviour in the icy mantle
compared with that required by evolution models
to match the heat flow from present-day Uranus is
beyond the scope of this paper.
However, our simulations are showing a
thermal boundary layer that might suppress convection and provide a
blanket to contain the heat in the central region of Uranus \citep{Stevenson1986,Podolak+Helled2012}.
This layer of impactor ice could also be
a compositional boundary if the icy material is not identical
to that in the proto-Uranus.
If these results can be usefully fed into evolution models,
then this could conceivably lead to another constraint on the types of impact
that are able to explain the current Uranus's thermal state
and perhaps also its unusual magnetic field.

\subsection{Orbiting Debris Field}

If the moons of Uranus are to form from the debris from the collision,
then it is necessary to place some rock into orbit beyond the Roche radius.
Satellites would also have to form beyond
than the co-rotation radius of $\sim$13~$R_\oplus$
to not have their orbits decay.
Using this instead of the Roche radius for our analysis
reduces the amount of material available by a few tens of percent
but does not change the overall conclusions.
As noted by S92, this task would be made easier by having less
differentiated bodies in the first place.
Nevertheless, for the higher angular
momentum collisions, our simulations succeed in placing
significant amounts of rock and ice into the debris field.
These clouds of debris are typically quite spherical
rather than disk-shaped,
with minimum-to-maximum axis ratios between 0.7 and 1,
defined using the inertia tensor.

\begin{figure}[t]
	\centering
	\plotone{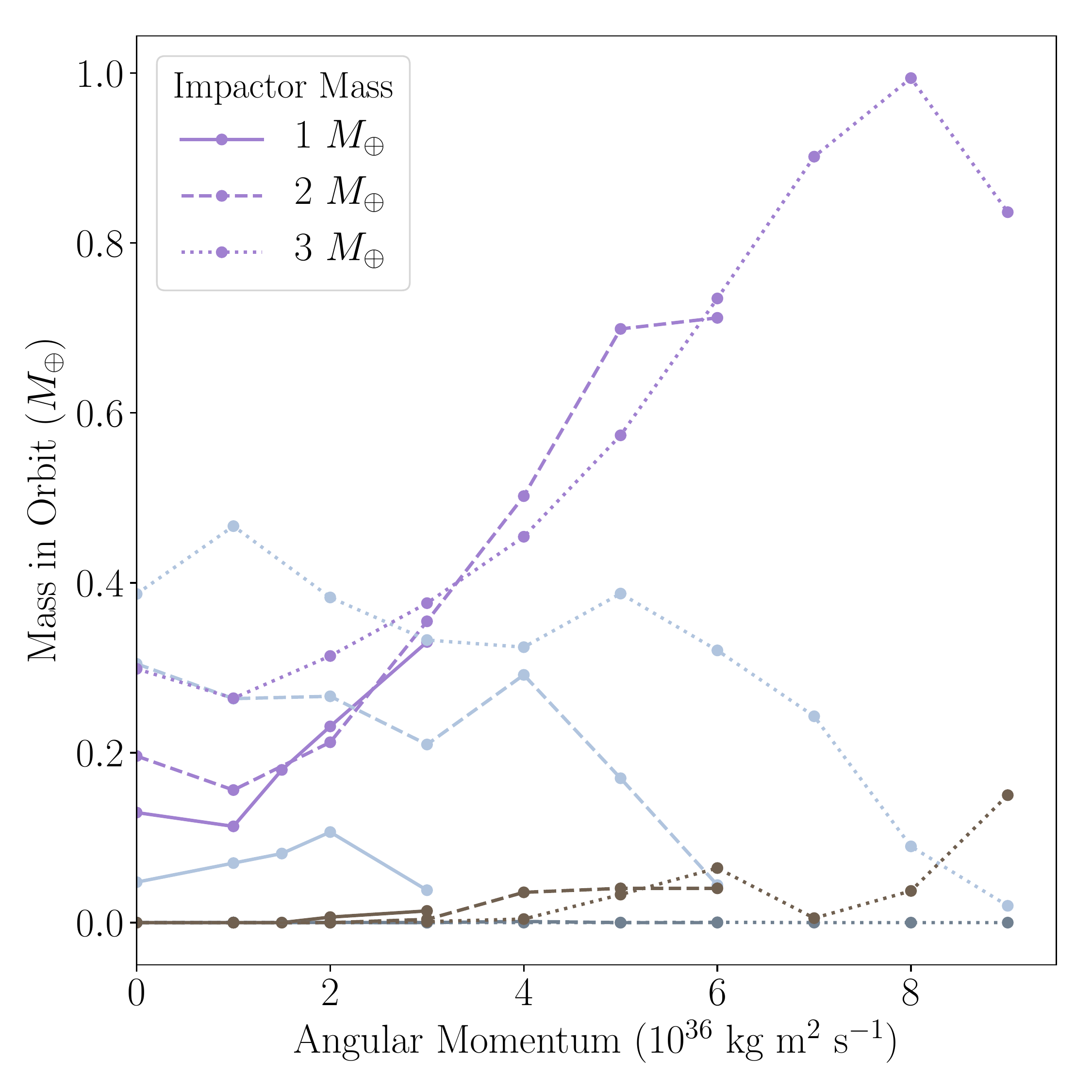}
	\caption{The masses of impactor and proto-Uranus material that are placed
		into orbit around the final planet
		(i.e. bound but outside a Roche radius of 6~$R_\oplus$)
		as functions of impactor mass and angular momentum.
		The line styles refer to the impactor mass and the
		colours to the material.
		Light and dark grey show the target's ice and rock material, respectively,
		and purple and brown show the same for the impactor.
		\label{fig:orbit_masses}}
\end{figure}

The amounts of rock and ice from the impactor and the proto-Uranus in
the debris cloud are shown in Figure~\ref{fig:orbit_masses},
as functions of impactor mass and angular momentum.
This shows how the more head-on collisions send more proto-Uranus ice into
orbit than impactor material.
The crossover to impactor ice being more prevalent in orbit
occurs at $L\approx3\times 10^{36}$~kg~m$^2$~s$^{-1}$
for impactors of mass 2 or 3~$M_\oplus$.
The lowest mass impactor never manages to eject more
proto-Uranus ice into orbit than impactor ice.

Grazing impacts sometimes involve multiple significant
collisions or near-miss passes,
creating large tidal streams of impactor material (Figure~\ref{fig:m2L5v5}).
For a more massive impactor
(and a correspondingly less massive proto-Uranus)
the impactor's core becomes less susceptible to tidal
stripping. Consequently, the higher mass impactors become less
efficient at placing rock into orbit in this way. It may be that
$>$3~$M_\oplus$ impactors would be too massive to leave any rock in
orbit via this mechanism.
These findings are broadly similar to those of S92;
though, they were restricted to $<$25 rock particles in orbit and,
for their more massive impactor cores,
found that only $<$3~$M_\oplus$ impactors could be disrupted enough
to leave rock in orbit.

\subsection{Atmosphere}

Most previous studies of atmospheric erosion during impacts have
focussed on vertical impacts onto terrestrial planets, where the
atmosphere comprises a much smaller mass fraction than is present in
our Uranus simulations \citep{Ahrens1993}. \citet{Shuvalov2009} performed
hydrodynamical simulations of oblique collisions of relatively small
projectiles (with sizes similar to the atmosphere's height)
into the Earth, finding more atmospheric erosion with more
oblique impacts. For sufficiently oblique impacts, the atmospheric
loss rose to all the mass above the horizon as seen from the point of impact.

For atmospheric erosion by giant impacts,
\citet{Genda+Abe2003} and \citet{Schlichting+2015} used a mixture of
analytical techniques and one-dimensional numerical simulations to
predict that the most important factor is the speed at which the
sub-atmospheric surface moves as a result of the shock wave
propagating through the planet.
This topic has also been little simulated in three dimensions.
\citet{Liu+2015b} tested, to our knowledge, the only previous
three-dimensional full-planet models,
with two simulations of head-on collisions on super-Earths.
The simulations presented here are the first in three dimensions
to quantify atmospheric erosion from giant impacts with inter-particle self-gravity
as well as the first to test a range of impact angles,
leaving much of this topic's huge parameter space still to be explored.

\begin{figure}[t]
	\centering
	\plotone{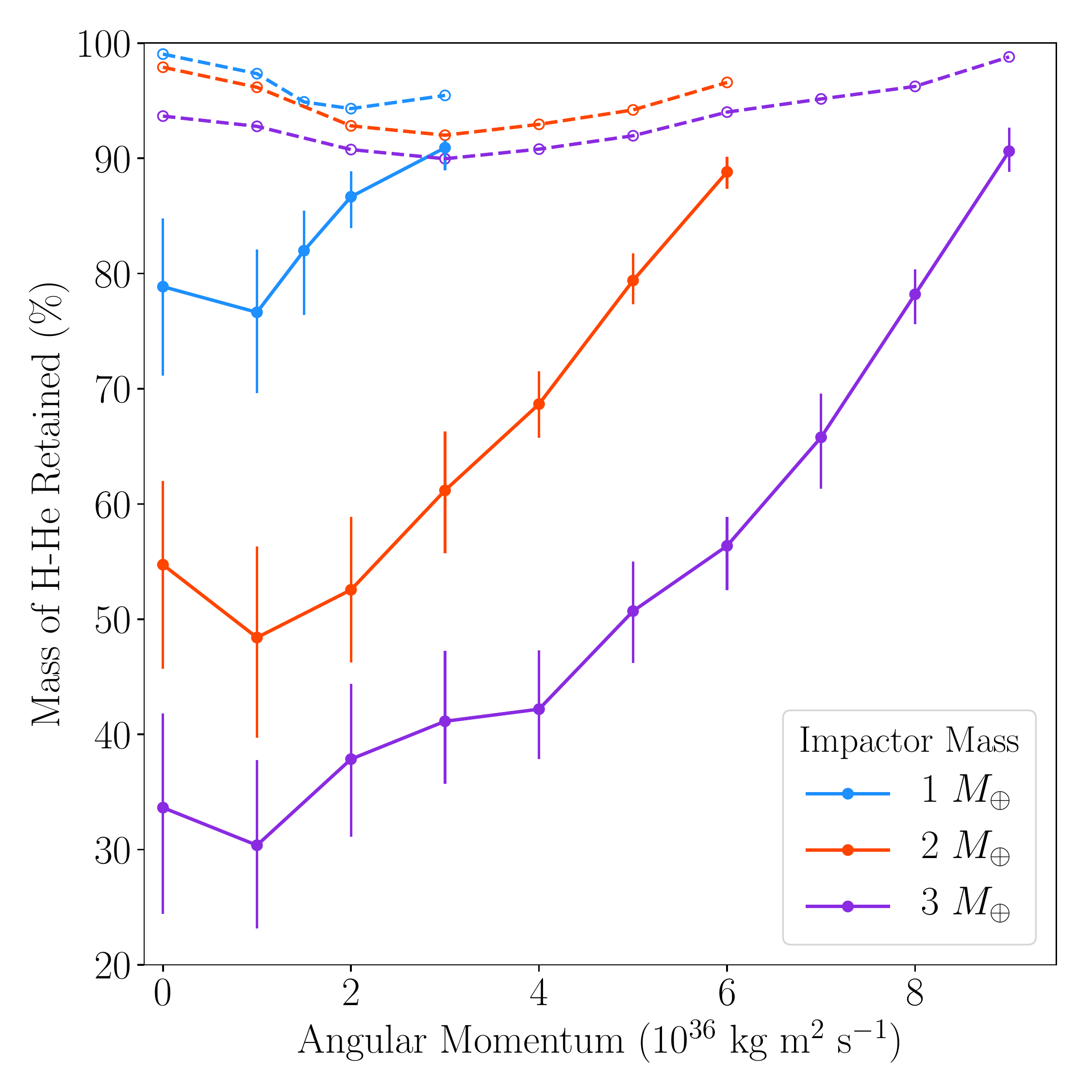}
	\caption{The mass fractions of the H--He atmosphere retained within
		a Roche radius of 6~$\pm$~0.5~$R_\oplus$ (solid
		lines) and still bound to the final planet (dashed
		lines), as functions of impactor mass and angular momentum.
		\label{fig:HHe_retention}}
\end{figure}

The fractions of the H--He atmosphere that are retained within the Roche
radius or bound to the final planet
following these giant impacts are shown in Figure~\ref{fig:HHe_retention},
as a function of impactor mass and angular momentum. Most of the eroded
atmosphere remains bound but can be jettisoned to large radii. There is a monotonic
behaviour with larger impactors eroding more atmosphere than smaller
ones, but the angular momentum dependence is more complicated.
The head-on collisions retain a few more per cent of the atmosphere within the
Roche radius than those with $L=1\times10^{36}$~kg~m$^2$~s$^{-1}$.
Up to half of the atmosphere can be
sent beyond the Roche radius for 2~$M_\oplus$ impactors,
and this rises to 70\% for $M_{\rm{i}}=3$~$M_\oplus$.

The proportion of the proto-Uranus H--He atmosphere that remains bound to
the final planet is always at least $\sim$90\%, with this minimum value
being reached for intermediate values of angular momentum at
$\sim$$3\times 10^{36}$~kg~m$^2$~s$^{-1}$.
More-grazing impacts lead to
significantly higher atmospheric retention
because not all the impactor's energy may be deposited at once,
especially if they undergo tidal stripping and multiple less-violent collisions.
As such, higher angular momentum giant impacts are less
effective at eroding the atmosphere, in contrast with the trends
determined by \citet{Shuvalov2009} for the different regime of much smaller impactors.
The atmosphere that is ejected by the giant impacts
typically originates from near to the impact site,
especially in the high angular momentum cases.
For the more head-on collisions, some atmosphere can also be lost
on the opposite side of proto-Uranus from where the impact occurs,
from the high outward velocities of the icy mantle.

\newpage
\section{Conclusions}\label{sec:conclusions}

We have performed SPH simulations to test the hypothesis that Uranus
endured a giant impact toward the end of its formation
and to investigate the consequences of such an event.
Animations of the simulated impacts are available at
\href{http://icc.dur.ac.uk/giant_impacts}{icc.dur.ac.uk/giant\_impacts}.
We confirm the findings of S92 that the impactor needs to have a mass of greater than
1~$M_\oplus$ in order to impart sufficient angular momentum to account for
Uranus' present rotation.

We also investigate where the impactor's mass
and energy are deposited within the planet. Sub-adiabatic temperature
gradients are typically created toward the outer regions of the icy
mantle, where most of the impactor ice is deposited. Higher impact
parameters can even lead to a temperature inversion near the top of the ice layer.
These more-grazing collisions also leave the impactor ice further out,
in a thin shell near the edge of the icy mantle,
whereas $\sim$head-on impacts can implant significant ice up to
0.5~$R_\oplus$ further inward and less-isotropically about the centre.
These findings may have important implications for understanding
the current heat flow (or rather the lack thereof) from Uranus' interior to its surface.

With our higher resolution simulations, we see significant
inhomogeneities in the deposited impactor material,
and can also properly resolve the composition of the debris field.
The impactor's ice can be quite isotropically distributed,
unlike its rocky core.
While most of this rock tends to end up at the top of the core
of the final planet, some small chunks become
embedded within its icy mantle. For higher angular
momentum impacts, significant amounts of rock and ice can be placed
into orbit during tidal disruption of the impactor. The efficiency of
this process is lower for 3 than 2~$M_\oplus$ impactors,
since the larger impactors are more able to resist tidal stripping,
but could still conceivably provide sufficient material to form
Uranus' current satellites if the angular momentum of the
collision exceeds $2\times10^{36}$~kg~m$^2$~s$^{-1}$.

While less than $\sim$10\% of the H--He atmosphere of the proto-Uranus
becomes unbound during the collisions, over half can be ejected to beyond
the Roche radius. This atmospheric erosion occurs more in lower
angular momentum collisions, where the impactor's energy is deposited
all at once and some atmosphere is also lost from the antipode to the impact point.

Higher numerical resolution simulations have allowed us to study
a variety of facets of the giant impact hypothesis
for producing Uranus' obliquity in detail,
including the first three-dimensional tests of atmospheric loss
with inter-particle self-gravity and from off-axis giant impacts.
Further work is under way to vary uncertain aspects such as
the material EoS and to increase the numerical resolution further
with a view to producing an improved prediction
for the internal structure and inhomogeneities in the final planet,
as well as testing the atmospheric erosion models of
\citet{Genda+Abe2003} and \citet{Schlichting+2015}.
The full particle data from the simulations are available
on reasonable requests for related studies or collaboration.

%
%
%
%

\acknowledgments
\section*{Acknowledgements}
We thank Lydia Heck and James Willis for their assistance with computational challenges.
We thank the anonymous referee for their helpful comments.
This work was supported by the Science and Technology Facilities Council (STFC)
grants ST/P000541/1 and ST/L00075X/1,
and used the DiRAC Data Centric system at Durham University,
operated by the Institute for Computational Cosmology on behalf of the
STFC DiRAC HPC Facility (www.dirac.ac.uk).
This equipment was funded by BIS National E-infrastructure capital grant ST/K00042X/1,
STFC capital grants ST/H008519/1 and ST/K00087X/1,
STFC DiRAC Operations grant ST/K003267/1 and Durham University.
DiRAC is part of the National E-Infrastructure.
J.A.K. is funded by STFC grant ST/N50404X/1.
V.R.E. acknowledges support from STFC grant ST/P000541/1.
R.J.M. acknowledges the support of a Royal Society University Research Fellowship.
L.F.A.T. and D.G.K. acknowledge support from NASA Outer Planets Research program award NNX13AK99G.
L.F.A.T., D.C.C., D.G.K., and K.J.Z. acknowledge support from NASA Planetary Atmospheres grant NNX14AJ45G.
C.L.F. was funded in part under the auspices of the U.S. Dept. of Energy,
and supported by its contract W-7405-ENG-36 to Los Alamos National Laboratory.



\appendix

\section{Equations of State}
\label{sec:app:eos}

\citet{Hubbard+MacFarlane1980} (HM80)'s equations of state
are expressed in terms of the temperature, $T$,
and density, $\rho$.
However, the simulation code uses the specific internal energy, $u$,
so we must convert between the two.
Including the energy contribution from the density:
\begin{align}
u_0(\rho)	&= \int_{\rho_0}^\rho \dfrac{P_0(\rho)}{\rho^2} \, d\rho \;, \\
u(\rho, T) &= u_0(\rho) + C_V T \;,
\end{align}
where $u_0(\rho)$ and $P_0(\rho)$ are
the specific internal energy and pressure at zero temperature,
$\rho_0$ is the material's zero-pressure density,
and $C_V$ is the specific heat capacity.

Using HM80's equations of state and expressions for $C_V$ and $P_0$,
the total pressure can then be tabulated
as a function of log($\rho$) and log($u$)
for interpolation in the SPH code.

HM80 did not provide expressions for the sound speed, $c_{\rm s}$,
so for simplicity we treat the H--He as an ideal gas
($c_{\rm s} = \sqrt{\gamma P / \rho\,}$)
and use approximate bulk moduli for the other materials:
$c_{\rm s} = \sqrt{2\times10^{10}~\rm{dyn~cm}^{-2} / \rho\,}$ for the ice mix,
and $c_{\rm s} = \sqrt{2\times10^{11}~\rm{dyn~cm}^{-2} / \rho\,}$ for the rocky core,
with the density in g~cm$^{-3}$ \citep{Matsui1996}.

\section{Impact Initial Conditions}
\label{sec:app:init_cond}

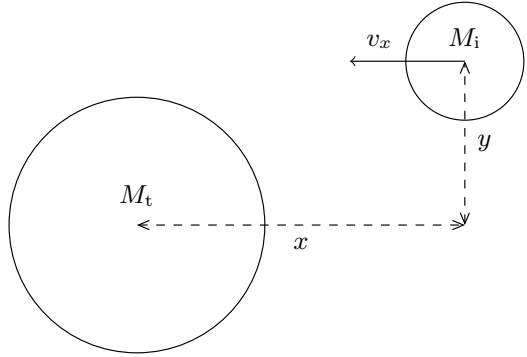
\begin{figure}[t]
	\centering
	\resizebox{0.4\textwidth}{!}{
		\begin{tikzpicture}[scale=0.4]
		\draw (0, 0) circle (3.9);
		\draw (10, 5) circle (1.8);
		\draw[>=angle 45, <->, dashed] (0, 0) -- (10, 0);
		\draw[>=angle 45, <->, dashed] (10, 0) -- (10, 5);
		\draw[->] (10, 5) -- ++(-3.5, 0);
		\node[below] at (5, -0.1) {$x$};
		\node[right] at (10.1, 2.5) {$y$};
		\node[above] at (7.4, 5.1) {$v_x$};
		\node[above] at (0, 0.3) {$M_{\rm{t}}$};
		\node[above] at (10, 5.1) {$M_{\rm{i}}$};
		\end{tikzpicture}
	}
	\caption{The relevant parameters for setting up an impact simulation:
		the $x$ and $y$ positions of the impactor,
		its initial velocity, $v_x$,
		and the masses of the proto-Uranus target (t) and impactor (i).
		\label{fig:impact_init_cond}}
\end{figure}

Figure~\ref{fig:impact_init_cond} shows the relevant input parameters
for an impact simulation in the target planet's rest frame.
The chosen parameters for each simulation are the impactor mass, $M_{\rm{i}}$
(and hence target mass, ${M_{\rm{t}} = 14.536~M_\oplus - M_{\rm{i}}}$),
total angular momentum, $L$,
and a velocity at infinity, $v_\infty=5$~km~s$^{-1}$.
From these inputs, we calculate the initial positions and velocities
of the two bodies.

In the centre-of-mass and zero-momentum frame,
the total angular momentum is
\begin{align}
L = L_z	&= M_{\rm{i}} \, v_{x,{\rm{i}}} \, y_{\rm{i}} + M_{\rm{t}} \, v_{x,\rm{t}} \, y_{\rm{t}} \nonumber \\
&= v_x \, y \left(M_{\rm{i}} \left(1-m'\right)^2 + M_{\rm{t}} \, m'^2\right) \;,
\label{eqn:ang_mom}
\end{align}
where  $m' \equiv M_{\rm{i}} / (M_{\rm{i}}+M_{\rm{t}})$.

In order to allow the bodies to be distorted tidally before the impact,
we set the initial separation $d$ of the two bodies such that,
at the point on the surface of the impactor that is closest to the target,
the gravitational force from the target planet is 10 times smaller
than that from the impactor:
\begin{equation}
d = \sqrt{\dfrac{10 \, M_{\rm{t}} R_{\rm{i}}^2}{M_{\rm{i}}}} \;.
\end{equation}
From conservation of energy, the velocity at a distance $d$ is
\begin{equation}
v_x = \sqrt{v_\infty^2 + \dfrac{2GM_{\rm{t}}}{d}} \;.
\end{equation}
Finally, $y$ is set using the chosen angular momentum
with Equation~\ref{eqn:ang_mom},
and $x = \sqrt{d^2 - y^2}$.

\end{document}